\documentclass[twocolumn,amsmath,amssymb,aps]{revtex4}

\usepackage{graphicx}
\usepackage{dcolumn}
\usepackage{bm}
\usepackage{textcomp}
\usepackage{url}
\DeclareGraphicsExtensions{.jpg,.eps,.pnm}

\begin{document}
\title{A higher-order spatial FDTD scheme with CFS PML \\ for 3D numerical simulation of wave propagation in cold plasma} \vspace{0.15cm}
\author{Konstantinos P. Prokopidis}

\affiliation{\vspace{0.15cm} Department of Electrical and Computer Engineering, Aristotle University of Thessaloniki, Thessaloniki, GR-54124, Greece. E-mail: kprokopi@ee.auth.gr  \vspace{0.15cm}}
\date{\today}

\begin{abstract}
A novel 3-D higher-order finite-difference time-domain framework with complex frequency-shifted perfectly matched layer for
the modeling of wave propagation in cold plasma is presented. Second- and fourth-order spatial approximations are
used to discretize Maxwell's curl equations and a uniaxial
perfectly matched layer with the complex frequency-shifted
equations is introduced to terminate the computational
domain. A numerical dispersion study of second- and higher-order techniques is elaborated and their stability criteria
are extracted for each scheme. Comparisons with analytical solutions verify the accuracy
of the proposed methods and the low dispersion error of the
higher-order schemes.
\end{abstract}

\maketitle

\section{Introduction}
The finite-difference time-domain (FDTD) method
\cite{Taflove2005,Inan_Marshall2011} is one of the most successful techniques in the
area of computational electromagnetics and has been widely used in
field computations in plasma and other dispersive media. Among the
FDTD techniques for dispersive media are included the recursive
convolution method, the $z$-transform method, the exponential
differencing technique, the auxiliary differential equation (ADE)
and integration equation methods. An extensive survey of the
previous techniques can be found in \cite{Cummer1997,Young2001}. The JE
convolution (JEC) method for plasma has also been introduced
\cite{Chen1998}.

Higher-order (HO) FDTD techniques \cite{Georgakopoulos2002} appeared
as a promising approach for the reduction of the inherent
numerical dispersion error of the conventional Yee scheme.
The HO FDTD algorithms have been extensively used in computational electromagnetics and have been optimized to
even reduce dispersion errors \cite{Zygiridis2007}, coupled with classical FDTD method on subgrids \cite{Georgakopoulos2001} and combined
with the alternative direction implicit (ADI) FDTD method \cite{Chen2002}.
Even though, HO methods have been widely used in many problems
involving simple dielectrics and waveguide problems \cite{Hwang2006}, they have not been applied in
dispersive media until recently \cite{Young1996,Prokopidis2004,Li2004,Fujii2004,Prokopidis2006,Bokil2012}. Although several HO
implementations for dispersive media have been proposed, they have not applied to three-dimensional problems. In this work, the proposed
formulation is tested in a three dimensional problem for which an analytical solution is available for comparison.

To complete the interior numerical scheme on a computational
domain, an absorbing boundary condition (ABC) should be used.
Although the original perfectly matched layer (PML) \cite{Taflove2005} is a
highly effective ABC, it can be applied, without modifications,
only to nondispersive media. Several PMLs has been extended to
handle plasma and general dispersive media \cite{Gedney1996,Fan2000,Prokopidis2008}
to name a few formulations.

In this work, a simple HO FDTD formulation with PML is presented
for the modeling of wave propagation in cold plasma. A study of stability is given for the
second order and the HO schemes. From error
analysis and numerical simulations, it is argued that HO
approaches provide higher accuracy than second-order schemes, with an expense of additional computations,
while maintaining the same memory requirements.

\section{FDTD formulation for cold plasma with higher order spatial approximations}
We assume unmagnetized cold plasma \cite{Inan2011} with
relative permittivity $\varepsilon_r(\omega)$ given by
\begin{equation}
\varepsilon_r(\omega)=1+\frac{\omega_p^2}{\omega (j \nu_c
-\omega)} \label{plasma}
\end{equation}
where $\omega_p$ is the radian plasma frequency, $\nu_c$ is the
collision frequency and assuming $e^{j \omega t}$ time dependence.
In the following we adopt the method of Young \cite{Young2001}, \cite{Werner_Mittra_2000}(Chapter 16) for cold plasma and we combine it with higher order spatial approximations.
The Amp\`{e}re's law in such a medium in a domain away from sources has the following form
\begin{equation}
\nabla \times \boldsymbol{\tilde{H}} = j \omega \varepsilon_0
\varepsilon_r (\omega) \boldsymbol{\tilde{E}} \label{Ampere_law}
\end{equation}
where the tilde denotes that the fields are in the frequency domain. We substitute (\ref{plasma}) into Amp\`{e}re's law and we get
\begin{equation}
\nabla \times \boldsymbol{\tilde{H}} = j \omega \varepsilon_{0}
\boldsymbol{\tilde{E}} + \boldsymbol{\tilde{J}}_p \label{Ampere_law_with_Jp}
\end{equation}
with the introduction of the variable $\boldsymbol{\tilde{J}}_p$ defined by
\begin{equation}
\boldsymbol{\tilde{J}}_p=\frac{\varepsilon_0 \omega_p^2}{j \omega +
\nu_c} \boldsymbol{\tilde{E}}. \label{Jp_def}
\end{equation}
The selection is such that $j \omega$ of (\ref{Ampere_law}) and of (\ref{plasma}) vanish each other.
We then transform (\ref{Ampere_law_with_Jp}) into the time domain
\begin{equation}
\nabla \times \boldsymbol{H} = \varepsilon_{0} \frac{d
\boldsymbol{E}}{dt} + \boldsymbol{J}_p
\end{equation}
After discretizing it at time $t=(n+1/2) \Delta
t$, using the central finite-difference and the central average operators with
respect to time given by $\delta_t f^n=f^{n+1/2}-f^{n-1/2}$
and $\mu_t f^{n}=(f^{n+1/2}+f^{n-1/2})/2$, we get the following equation
\begin{equation}
(\nabla \times \boldsymbol{H})^{n+1/2} = \varepsilon_{0}
\frac{\delta_t \boldsymbol{E}^{n+1/2}}{\Delta t} + \boldsymbol{J}_p^{n+1/2} \label{Ampere_dif}
\end{equation}
where it is assumed that the variable $\boldsymbol{J}_p$ is defined at the same time instance to the magnetic fields i.e. at $(n+1/2) \Delta t$.
The update equation for $\boldsymbol{E}$ is obtained
\begin{equation}
\boldsymbol{E}^{n+1} =\boldsymbol{E}^{n}+\frac{\Delta
t}{\varepsilon_0}(\nabla \times
\boldsymbol{H})^{n+1/2} - \frac{\Delta t}{\varepsilon_0} \boldsymbol{J}_p^{n+1/2}
\end{equation}
Transforming (\ref{Jp_def}) into the time domain, we obtain the
first-order differential equation for variable $\boldsymbol{J}_p$
\begin{equation}
\frac{d \boldsymbol{J}_p}{dt}+\nu_c \boldsymbol{J}_p=\varepsilon_0
\omega_p^2 \boldsymbol{E}
\end{equation}
We write the previous equation in operational form at the time step $t=n \Delta
t$
\begin{equation}
\frac{\delta_t \boldsymbol{J}_p^n}{\Delta t}+\nu_c \mu_t
\boldsymbol{J}_p^n=\varepsilon_0 \omega_p^2 \boldsymbol{E}^n \label{Jp_dif}
\end{equation}
and the update equation for variable $\boldsymbol{J}_p$ is the
following
\begin{equation}
\boldsymbol{J}_p^{n+1/2} =\frac{2-\nu_c \Delta t}{2 + \nu_c \Delta
t}\boldsymbol{J}_p^{n-1/2} +\frac{2 \varepsilon_0 \omega_p^2 \Delta
t}{2  + \nu_c \Delta t} \boldsymbol{E}^{n}
\end{equation}
The proposed FDTD scheme is different than that of \cite{Taflove2005} in the fact that is is not semi-implicit. The proposed scheme
uses one additional variable (the variable $J_p$) as the JEC scheme \cite{Chen1998}, but it is better in terms of memory requirements than the direct $D-E$
implementation, based in the differential equation as exposed in \cite{Prokopidis2004}.

As with the standard FDTD scheme, the temporal derivatives of the
proposed method are discretized using second order approximations.
On the contrary, the central spatial operator of $N$-order ($N$:
even number)
\begin{equation}
\left( \frac{\partial f}{\partial \beta} \right)^{n \Delta t}_{m
\Delta \beta} \approx \frac{1}{\Delta \beta} \sum_{l=1, (l \,
\textrm{odd})}^{N-1}c_{l}^{N}\left( f^n_{m+l/2}-f^n_{m-l/2}\right)
\end{equation}
is invoked for the spatial derivatives, where the coefficients
$c_{l}^{N}$ are given by an analytical expression
\cite{Prokopidis2004}, e.g., $c_{1}^{2}=1$ (Yee scheme or (2,2) scheme),
$c_{1}^{4}=9/8$ and $c_{3}^{4}=-1/24$ (fourth-order scheme or (2,4) scheme) and
$c_{1}^{6}=75/64$, $c_{3}^{6}=-25/384$ and $c_{5}^{6}=3/640$ (sixth-order scheme or (2,6) scheme).

\begin{figure}[h]
\centering
\includegraphics[width=9cm]{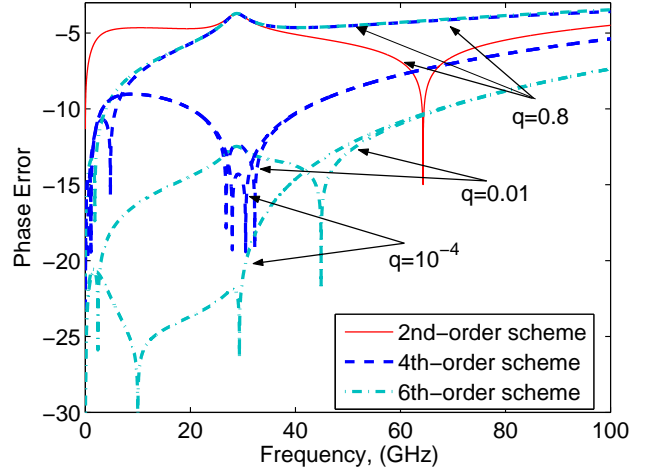}
\caption{Phase error introduced by the second-, fourth-, and
sixth-order schemes with various values of the Courant number
$q$.} \label{fig1}
\end{figure}

\begin{figure}[h]
\centering
\includegraphics[width=9cm]{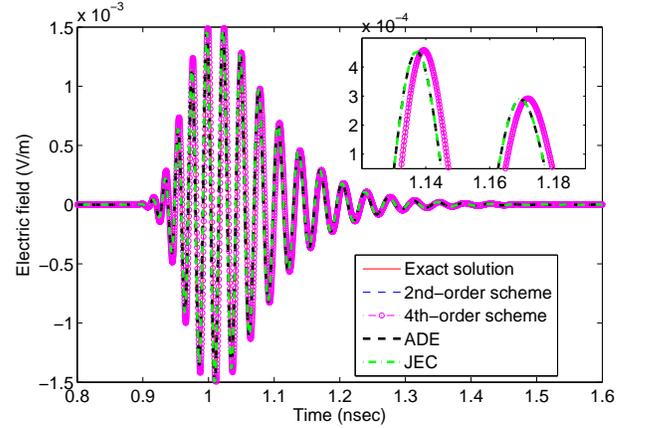}
\caption{Electric field computed using the exact solution,
second-order scheme (Young's technique \cite{Young2001}), proposed fourth-order scheme, ADE \cite{Prokopidis2004} and JEC \cite{Chen1998} methods at distance $0.25$ m from the source. We observe that the solution of the fourth-order scheme coincide with the analytical one.} \label{fig2}
\end{figure}

\begin{figure}[h]
\centering
\includegraphics[width=9cm]{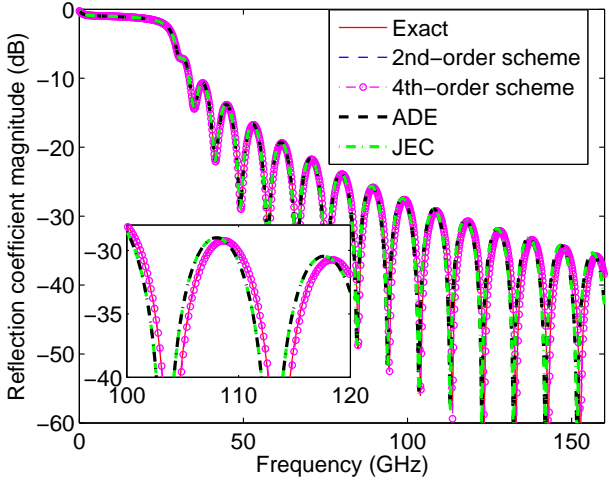}
\caption{Reflection coefficient magnitude computed using the
second-order scheme (Young's technique \cite{Young2001}), proposed fourth-order scheme, ADE \cite{Prokopidis2004} and JEC \cite{Chen1998} methods in
comparison to the analytical solution. We observe that the solution of the fourth-order scheme coincide with the analytical one.} \label{fig3}
\end{figure}

\section{Numerical Dispersion Errors and Stability Analysis}
\subsection{Numerical dispersion analysis}
To consider the dispersion errors created by the aforementioned
scheme, we assume that the fields
$\boldsymbol{\tilde{E}},\boldsymbol{\tilde{H}}, \boldsymbol{\tilde{J}}_p$ are plane waves of
the form $\boldsymbol{\tilde{F}}=\boldsymbol{\tilde{F}}_0 \exp(j
\omega t - j \boldsymbol{k} \boldsymbol{r} )$. Then
in the discretized space the fields are
\begin{equation}
\boldsymbol{\tilde{F}}^n=\boldsymbol{\tilde{F}}_0 \exp(j \omega n \Delta t - j (\tilde{k}_x I \Delta x + \tilde{k}_y J \Delta y + \tilde{k}_z K \Delta z))
\end{equation}
where $\tilde{\boldsymbol{k}}$ is the numerical wavenumber given by $\tilde{\boldsymbol{k}} = \tilde{k}_x \boldsymbol{a}_{x} + \tilde{k}_y \boldsymbol{a}_{y} + \tilde{k}_z \boldsymbol{a}_{z}$ and indexes $I, J, K$ denote the position of the nodes in the FDTD mesh. Equations (\ref{Ampere_dif}) and (\ref{Jp_dif}) are written in discretized form
using the following definitions \cite{Prokopidis2004}
\begin{equation}
\nabla \leftarrow - j \tilde{\boldsymbol{k}}, \, \mu_t \leftarrow \cos(\omega \Delta t /2), \, \delta_t \leftarrow 2 j \sin(\omega \Delta t /2)
\end{equation}
as
\begin{gather}
j \tilde{\boldsymbol{k}} \times \tilde{\boldsymbol{H}_0}= \epsilon_0 \frac{2 j \sin(\omega \Delta t /2)}{\Delta t} \tilde{\boldsymbol{E}_0} + \tilde{\boldsymbol{J}}_{p0}  \\
\frac{2 j \sin(\omega \Delta t / 2)}{\Delta t} \tilde{\boldsymbol{J}}_{p0} + \nu_c \cos( \omega \Delta t /2) \tilde{\boldsymbol{J}_{p0}} = \epsilon_0 \omega_p^2 \tilde{\boldsymbol{E}_0}
\end{gather}
where the numerical wavenumber $\tilde{\boldsymbol{k}}$ is given by
\begin{equation}
\tilde{\boldsymbol{k}}=\sum_{\beta=x,y,z} \frac{2}{\Delta
\beta} \sum_{l=1,(l \, \textrm{odd})}^{N-1} c^N_l \sin \left (
\frac{l \tilde{k}_{\beta} \Delta \beta}{2} \right )
\boldsymbol{a}_{\beta}
\end{equation}
with $\boldsymbol{a}_{\beta}$ the unit vector in the $\beta$
direction.

For easy reference we give here the numerical wavenumber for the three most used schemes. For the (2,2) scheme
we have
\begin{equation}
\tilde{\boldsymbol{k}}=\sum_{\beta=x,y,z} \frac{2}{\Delta
\beta} \sin \left ( \frac{\tilde{k}_{\beta} \Delta \beta}{2} \right )
\boldsymbol{a}_{\beta}
\end{equation}
for the (2,4) scheme
\begin{equation}
\begin{split}
\tilde{\boldsymbol{k}} & =\sum_{\beta=x,y,z} \frac{2}{\Delta
\beta} \left[ \frac{9}{8} \sin \left (\frac{ \tilde{k}_{\beta} \Delta \beta}{2} \right ) \right. \\
& \left. - \frac{1}{24} \sin \left (\frac{3 \tilde{k}_{\beta} \Delta \beta}{2} \right ) \right ]
\boldsymbol{a}_{\beta}
\end{split}
\end{equation}
and for the (2,6) scheme
\begin{equation}
\begin{split}
\tilde{\boldsymbol{k}} & =\sum_{\beta=x,y,z} \frac{2}{\Delta
\beta} \left[ \frac{75}{64} \sin \left (\frac{ \tilde{k}_{\beta} \Delta \beta}{2} \right ) \right. \\
& \left. - \frac{25}{384} \sin \left (\frac{3 \tilde{k}_{\beta} \Delta \beta}{2} \right ) + \frac{3}{640} \sin \left (\frac{5 \tilde{k}_{\beta} \Delta \beta}{2} \right )\right ]
\boldsymbol{a}_{\beta}
\end{split}
\end{equation}

These equations are combined in a single equation
\begin{gather}
-j \tilde{\boldsymbol{k}} \times \tilde{\boldsymbol{H}_0} = \\
j \tilde{\omega} \varepsilon_0 \left( 1+ \frac{\omega_p^2}{j \tilde{\omega} (j \nu_c \cos(\omega \Delta t/2) +  \tilde{\omega})   } \right) \tilde{\boldsymbol{E}_0}
\end{gather}
where $\tilde{\omega}=(2 / \Delta t) \sin (\omega \Delta t /2)$
which corresponds to Amp\`{e}re's law
\begin{equation}\label{Ampere_discrete}
-j \tilde{\boldsymbol{k}} \times \tilde{\boldsymbol{H}_0} = j \tilde{\omega} \varepsilon_{\textrm{num}} \tilde{\boldsymbol{E}_0}
\end{equation}
It can be easily deduced that the numerical permittivity for the
proposed scheme is
\begin{equation}
\tilde{\varepsilon} = \varepsilon_0 \left\{ 1 +
\frac{\omega_p^2}{ \tilde{\omega} [j \nu_c \cos(\omega \Delta
t / 2) - \tilde{\omega} ]}\right\} \label{num_permittivity}
\end{equation}
which is exactly the same as of the Young's scheme \cite{Young2001} as expected, but it is different to the ADE approach of \cite{Prokopidis2004}.

If the above discretized Amp\`{e}re's law \eqref{Ampere_discrete} is combined with the Faraday's law
\begin{equation}
 j \tilde{\omega} \mu_0 \tilde{\boldsymbol{H}_0} = j
\tilde{\boldsymbol{k}} \times \tilde{\boldsymbol{E}_0}
\end{equation}
the numerical dispersion relation is derived
\begin{equation}
j \tilde{\omega} \tilde{\varepsilon} j \tilde{\omega} \mu_0 = -
\tilde{\boldsymbol{k}} \cdot \tilde{\boldsymbol{k}}.
\label{num_dispersion}
\end{equation}

To bring to light some salient features of the proposed scheme, we
define the phase error as $e_{\textrm{phase}}=\ln | \Re e
\{k - \tilde{k} \}/ \Re e \{k \}|$, where $\Re e$ denotes the
real part, $\tilde{k}$ is a solution of
(\ref{num_dispersion}) and $k$ is a solution of the dispersion
relation of the continuous space. We consider the plasma model
with parameters: $\omega_p=2 \pi \, 28.7 \times 10^9$ rad/sec,
$\nu_c=20 \times 10^9$ rad/sec, and FDTD parameters: $\Delta
x=0.5$ mm with different values of time step $\Delta t=q \Delta x
/ c_0$, where $q$ is the Courant number and $c_0$ is the velocity
of light in vacuum. In Fig. \ref{fig1} a comparison of second-,
fourth-, and sixth-order schemes is presented in terms of the
dispersion error they introduce. It is observed that since the HO
schemes are second-order accurate in time we should take very
small time step in order to improve temporal accuracy and improve the accuracy of the overall scheme.
Moreover, a very small time step for the second-order scheme (with fixed cell size) cannot improve the accuracy of the overall method,
since the error from the rough approximation of the spatial derivatives contaminate the solution. 

\subsection{Stability analysis}
The stability condition of the proposed formulation can be derived by a combination of the von Neumann method and Routh-Hurwitz criterion \cite{Pereda2001}. In fact, if we transform the numerical dispersion relation \eqref{num_dispersion} to the $z$-domain ($z=e^{j \omega \Delta t}$) we get the stability polynomial
\begin{equation}
\frac{(z-1)^2}{z} \tilde{\varepsilon}_r(z) + (c_0 \Delta t)^2 \tilde{\boldsymbol{k}} \cdot \tilde{\boldsymbol{k}} =0
\label{stability_polynomial}
\end{equation}
where the numerical relative permittivity $\tilde{\varepsilon}_r(z)$ is given from (\ref{num_permittivity}) after some algebra, by
\begin{equation}
\tilde{\varepsilon}_r (z)=1 + \frac{(\omega_p \Delta t)^2}{0.5 \nu_c \Delta t (z-z^{-1}) + (z-2+z^{-1})}
\end{equation}

It can be concluded from \eqref{stability_polynomial} that the stability polynomial
is
\begin{equation}
(z-1)^2 \tilde{\varepsilon}(z) + 4 z \nu^2 =0
\end{equation}
where $\nu^2$ is given by
\begin{equation}
\nu^2=(c_0 \Delta t)^2 \sum_{\beta=x,y,z} \frac{1}{\Delta
\beta ^2} \left[ \sum_{l=1,(l \, \textrm{odd})}^{N-1} c^N_l \sin \left (
\frac{l \tilde{k}_{\beta} \Delta \beta}{2} \right ) \right]^2
\end{equation}
After algebraic manipulations and the application of the bilinear transform $z=(r+1)/(r-1)$, we get the following stability polynomial with respect to $r$
\begin{equation}
\begin{split}
S(r)=& 2 \nu^2 \nu_c \Delta t \, r^3 +
(\omega_p^2 \Delta t^2 + 4 \nu^2) r^2+ \\
& 2 \nu_c \Delta t (1-\nu^2) r + 4-4 \nu^2 - \omega_p^2 \Delta t^2
\end{split}
\end{equation}
with the corresponding Routh table shown in Table \ref{Routh_table}. In order the scheme to be stable
the values of the first row of the Routh table should be non-negative quantities. After some algebra, we have
\begin{equation}
c_3=\frac{2 \nu_c \omega_p^2 \Delta t^3}{4 \nu^2 + \omega_p^2 \Delta t^2}, \,\, c_4=2 \nu_c \Delta t (1- \nu^2)
\end{equation}
We impose the inequalities and we get the following restrictions
\begin{equation}
\nu^2 \leq 1, \nu_c \geq 0
\end{equation}
It can be easily proved for the (2,2) scheme that the relation $\nu^2 \leq 1$ leads to the conventional FDTD stability criterion. For the case
of the (2,4) scheme we have
\begin{equation}
\begin{split}
\Delta t \leq & \frac{1}{c_0} \left[ \sum_{\beta=x,y,z}  \frac{1}{\Delta \beta^2} \left( \frac{9}{8} \sin \left( \frac{\tilde{k}_{\beta} \Delta \beta}{2} \right) \right. \right. \\
& \left. \left. - \frac{1}{24} \sin \left( \frac{3 \tilde{k}_{\beta} \Delta \beta}{2} \right) \right)^2 \right]^{-1/2}
\end{split}
\end{equation}
For practical cases, the worst case is considered, where both $\sin()$ take such values in order the quantity in brackets to be maximum (and the stability criterion is the most restrictive). As a result of the previous assumption the stability criterion for the (2,4) case is
\begin{equation}
\Delta t \leq \frac{6}{7 \, c_0} \left( \sum_{\beta=x,y,z}  \frac{1}{\Delta \beta^2} \right)^{-1/2}
\end{equation}
In a similar manner, the stability condition for the (2,6) case yields
\begin{equation}
\Delta t \leq \frac{120}{149 \, c_0} \left( \sum_{\beta=x,y,z}  \frac{1}{\Delta \beta^2} \right)^{-1/2}
\end{equation}

\begin{table}[t]
\caption{Routh table}
\centering
\begin{tabular}{c c}
\hline
$2 \nu^2 \nu_c \Delta t$ & $2 \nu_c \Delta t (1 - \nu^2)$ \\
$4 \nu^2 + \omega_p^2 \Delta t^2$ & $4-4 \nu^2 - \omega_p^2 \Delta t^2$ \\
$c_3$ & $0$ \\
$c_4$ & $0$ \\
\hline
\end{tabular}
\label{Routh_table}
\end{table}
It is observed that the stability criterion of the HO schemes is stricter than the second-order scheme--a remark useful for practical simulations.

\section{CFS-PML formulation}

We assume that a PML terminates unmagnetized cold plasma with
relative permittivity $\varepsilon_r(\omega)$ given by (\ref{plasma}).
Following the uniaxial formulation of the PML (UPML), initially
introduced by Sacks \textit{et al.} \cite{Sacks1995} and adapting the complex frequency shifted
(CFS) approach proposed by Kuzuoglu \cite{Kuzuoglu1996}, we propose a PML formulation
for the case of the cold plasma that we will use it in HO FDTD grids. The proposed formulation
for the case of the second-order FDTD schemes has been already exposed in \cite{Prokopidis2008} but it is
included here for clarity.

The modified Maxwell's curl equations inside the PML region in the frequency domain can be written as
\begin{gather}
\nabla \times \boldsymbol{\tilde{H}} = j \omega \varepsilon_0
\varepsilon_r (\omega)\mathcal{T} \cdot \boldsymbol{\tilde{E}} \label{Ampere_law_PML}\\
\nabla \times \boldsymbol{\tilde{E}} = - j \omega \mu_0
\mathcal{T} \cdot \boldsymbol{\tilde{H}} \label{Faraday_law}
\end{gather}
where ${\mathcal{T}}$ is the diagonal ``material'' tensor defined
by ${\mathcal{T}}= \textrm{diag} \{\zeta_x/(\zeta_y
\zeta_z),\zeta_y/(\zeta_z \zeta_x),\zeta_z/(\zeta_x \zeta_y) \}$. The definition of the
stretching coefficients $\zeta_s$ is
\begin{equation}
\zeta_s=1/\left(\kappa_s+\frac{\sigma_s}{\gamma+j \omega}\right),
s=x,y,z \label{str_coef}
\end{equation}
where $\gamma$ is considered constant in this work. Amp\`{e}re's
law (\ref{Ampere_law_PML}) is written as
\begin{equation}
\nabla \times \boldsymbol{\tilde{H}} = j \omega \varepsilon_{0}
\mathcal{T} \cdot \boldsymbol{\tilde{E}} + \boldsymbol{\tilde{Q}}
\label{Ampere_Drude}
\end{equation}
with the introduction of the variable $\boldsymbol{\tilde{Q}}$
defined by
\begin{equation}
\boldsymbol{\tilde{Q}}=\frac{\varepsilon_0 \omega_p^2}{j \omega +
\nu_c} \mathcal{T} \cdot \boldsymbol{\tilde{E}} \label{Qe}
\end{equation}
We introduce the variable $\boldsymbol{\tilde{R}} = \mathcal{T}
\cdot \boldsymbol{\tilde{E}}$ and after transformation into the
time domain, (\ref{Ampere_Drude}) takes the form
\begin{equation}
\nabla \times \boldsymbol{H} = \varepsilon_{0} \frac{d
\boldsymbol{R}}{dt} + \boldsymbol{Q}
\label{Amprere_Drude}
\end{equation}
After discretizing (\ref{Amprere_Drude}) at time $t=(n+1/2) \Delta
t$, we get the following equation with operators
\begin{equation}
(\nabla \times \boldsymbol{H})^{n+1/2} = \varepsilon_{0}
\frac{\delta_t \boldsymbol{R}^{n+1/2}}{\Delta t} + \boldsymbol{Q}^{n+1/2}
\end{equation}
and the update equation for $\boldsymbol{R}$ is obtained
\begin{equation}
\boldsymbol{R}^{n+1} =\boldsymbol{R}^{n}+\frac{\Delta
t}{\varepsilon_0}(\nabla \times
\boldsymbol{H})^{n+1/2} - \frac{\Delta t}{\varepsilon_0} \boldsymbol{Q}^{n+1/2}
\end{equation}
Transforming (\ref{Qe}) into the time domain, we obtain the
first-order differential equation for variable $\boldsymbol{Q}$
\begin{equation}
\frac{d \boldsymbol{Q}}{dt}+\nu_c \boldsymbol{Q}=\varepsilon_0
\omega_p^2 \boldsymbol{R}
\end{equation}
We write the previous equation in operational form
\begin{equation}
\frac{\delta_t \boldsymbol{Q}^n}{\Delta t}+\nu_c \mu_t
\boldsymbol{Q}^n=\varepsilon_0 \omega_p^2 \boldsymbol{R}^n
\end{equation}
and the update equation for variable $\boldsymbol{Q}$ is the
following
\begin{equation}
\boldsymbol{Q}^{n+1/2} =\frac{2-\nu_c \Delta t}{2 + \nu_c \Delta
t}\boldsymbol{Q}^{n-1/2} +\frac{2 \varepsilon_0 \omega_p^2 \Delta
t}{2  + \nu_c \Delta t} \boldsymbol{R}^{n}
\end{equation}

From the definition of the variable $\boldsymbol{\tilde{R}}$, the
$x$ coordinate component is derived as $\tilde{R}_x = \zeta_x
/(\zeta_y \zeta_z) \tilde{E}_x$. Similarly, we define variable
$\boldsymbol{\tilde{S}}$ such that the $x$-component to be
$\tilde{S}_x = (\zeta_x / \zeta_y) \tilde{E}_x$. Thus, the
differential equations relating $S_x$ with $E_x$ and $R_x$ with
$S_x$ take the following form
\begin{gather}
\kappa_x \frac{d S_x}{dt}+(\kappa_x \gamma + \sigma_x)
S_x=\kappa_y \frac{d E_x}{dt}+(\kappa_y \gamma + \sigma_y) E_x \label{Ex_update}\\
\frac{d R_x}{dt}+\gamma R_x=\kappa_z \frac{d S_x}{dt}+(\kappa_z
\gamma+\sigma_z) S_x \label{Sx_update}
\end{gather}
The update equations of $E_x$ and $S_x$ are derived from
(\ref{Ex_update}), (\ref{Sx_update}) and similar equations can be
obtained for all the other components.

\section{Numerical Results}
To investigate the accuracy of the proposed scheme, we assume an
one-dimensional problem for which a closed-form solution is
available with the use of Fourier transforms. We consider the propagation of the Gaussian pulse
$g(t)=\exp\{-(t-5 \tau)^2 / (2 \tau^2)\}$, where $\tau=10.6 \times
10^{-12}$ sec in plasma with parameters: $\omega_p=2 \pi \, 20
\times 10^9$ rad/sec, $\nu_c=20 \times 10^9$ rad/sec. The spatial
discretization is $0.5$ mm, $q=0.9$ for the second-order schemes
and $q=0.1$ for the fourth-order scheme. Simulations were carried
out over $8000 \Delta t$ and $10000 \Delta t$ for the second- and
fourth-order schemes, respectively. In Fig. \ref{fig2} we compare
the FDTD results with the analytical solution for the electric
field waveform. It is observed that all the second-order schemes have identical accuracy, while
the proposed fourth-order order technique is very close to the analytical solution,
as shown in the inset of Fig. \ref{fig2}, indicating
its higher accuracy.

We next calculate the reflection coefficient of a plasma slab
($\omega_p=2 \pi \, 28.7 \times 10^9$ rad/sec, $\nu_c=20 \times
10^9$ rad/sec) with thickness $1.5$ cm. The spatial step is
$\Delta z =75 \mu$m and the time step is $\Delta t=0.5 \Delta z /
c_0$. In this simulation, we choose the same spatial and temporal
steps for the second- and fourth-order schemes. The computational
domain is subdivided into $12000$ cells and the simulation time is
$30000 \Delta t$. The one-sided approximations of \cite{Yefet2001}
were used in the boundaries for the fourth-order scheme. Fig.
\ref{fig3} shows the magnitude of the reflection coefficient
computed using the proposed second- and fourth-order schemes, the
ADE, JEC techniques and the analytical solution. The increased accuracy of the
HO scheme is clearly demonstrated.

To validate the proposed FDTD schemes and the introduced PML, we
compare the FDTD results with the analytical solution for a
three-dimensional problem. We consider the transient field
produced by an infinitesimal electric dipole in infinite
homogeneous plasma with parameters: $\omega_p=2 \pi \, 9 \times
10^9$ rad/sec, $\nu_c=2 \times 10^9$ rad/sec. The source is the
Gaussian pulse mentioned previously and is located at the origin.
The Amp\`{e}re's law in the time domain is modified as follows with the inclusion of
the source excitation $\boldsymbol{J}_i$
\begin{equation}
\nabla \times \boldsymbol{H} = \varepsilon_{0} \frac{d
\boldsymbol{E}}{dt} + \boldsymbol{J}_p + \boldsymbol{J}_i
\end{equation}
where $\boldsymbol{J}_i=\frac{I_0(t)}{\Delta x \Delta y} \boldsymbol{a}_z$ and $I_0(t)$ is the Gaussian pulse.

The analytical time-dependent solution is obtained through an
inverse fast Fourier transform (IFFT) of the frequency-domain
analytical solution as described in Appendix \ref{Ap_A}. The computational domain is divided by $50
\times 50 \times 50$ cubic cells with size $\Delta=0.5$ mm and the
time step is $\Delta t=q \Delta / (c_0 \sqrt{3})$, with $q=0.3$.
The PML is six cells thick, the conductivity $\sigma_s$ and the
parameter $\kappa_s$ of (\ref{str_coef}) are subject to
fourth-order polynomial scaling according to \cite{Gedney1996} with
$\sigma_{s,\textrm{max}}=4.602 \times 10^{12}$,
$\kappa_{s,\textrm{max}}=2$, and $\gamma=0.5$.

\begin{figure}[t]
\centering
\includegraphics[width=9cm]{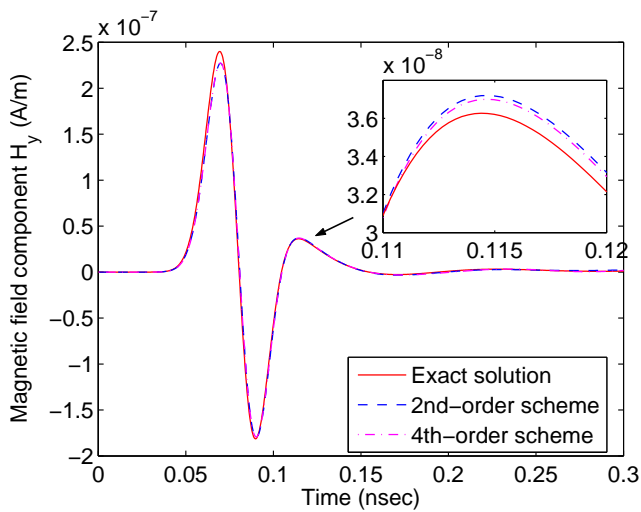}
\caption{Magnetic field component $H_y$ at distance $\boldsymbol
r=[16 \Delta,0,0]$ excited by an electric dipole at the origin in
homogeneous plasma, calculated analytically and using second- and
fourth-order FDTD schemes.} \label{fig4}
\end{figure}

\section{Conclusion}
In this work, we have introduced a HO FDTD formulation for wave
propagation in plasma and extended, for the first time, the efficient
CFS approach of the PML for HO FDTD methods in dispersive media. It was found that
the proposed method of second-order has the same accuracy as the
ADE and JEC techniques and possesses the same memory requirements
as the JEC method. Additionally, the introduced fourth-order
scheme exhibits lower dispersion error than the conventional
second-order schemes and it is very effective for long-time and/or
long-distance simulations.

\begin{acknowledgments}
The author would like to thank Dr Dimitrios C. Zografopoulos for helpful discussions.
\end{acknowledgments}

\appendix
\section{Analytical solution of infinitesimal electric dipole in plasma medium}\label{Ap_A}
We assume an infinitesimal electric dipole of length $\ell$ in air.
The magnetic field in the near-field of the dipole in air is given by \cite{Balanis1997}
\begin{equation}
H_{\phi}=j \frac{k I_0 \ell \sin \theta}{4 \pi r} \left( 1+ \frac{1}{j k r} \right) e^{-j k r}
\end{equation}
where $k$ is the wavenumber and $I_0$ is the current of the dipole, which is not a function of the space coordinates. If the surrounding
medium is a dispersive medium with relative dielectric permittivity $\varepsilon_r(\omega)$ instead of air, the wave number $k$ has the form $k=b- j a$
where $a$ and $b$ are the attenuation and the phase constants respectively
\begin{equation}
a=-\frac{\omega}{c_0} \Im m \{ \sqrt{\varepsilon_r(\omega)}\}, b=\frac{\omega}{c_0} \Re e \{ \sqrt{\varepsilon_r(\omega)}\}
\end{equation}
If the current $I_0$ is a time-dependent function, the magnetic field in the time domain is given by
\begin{equation}
H_{\phi}(t,r)= \mathcal{F}^{-1} \{H_{\phi}(\omega,r) I_0(\omega) \}
\end{equation}
where $I_0(\omega)$ is the Fourier transform of the excitation $I_0(t)$ and $\mathcal{F}^{-1}$ denotes the inverse Fourier transform.
It can be observed that for $\phi=0$ and $\theta=\pi/2$, i.e. at the $x$-axis $H_y=H_{\phi}$ -- a remark we exploited in order to compare the field values of
the rectangular FDTD grid with the analytical solution in spherical coordinates.

\section{FDTD equations at material interfaces}
We assume the case of an interface between air and plasma. Using the most common approximation of the averaging the dielectric permittivities of the two adjacent media, the permittivity at the interface is given by
\begin{equation}
\varepsilon_{\textrm{int}} (\omega) = \cfrac{1}{2} \left( \varepsilon_0 + \varepsilon_0 \varepsilon_r(\omega) \right)
\end{equation}
where $\varepsilon_r(\omega)$ is given by \eqref{plasma}. It can be easily concluded that the Amp\`{e}re's law remains unchanged with the definition
of the $\boldsymbol{\tilde{J}}_p$ as
\begin{equation}
\boldsymbol{\tilde{J}}_p=\frac{\varepsilon_0 \omega_p^2}{2 (j \omega +
\nu_c)} \boldsymbol{\tilde{E}}
\end{equation}

\bibliographystyle{nature}
\bibliography{ReferenceDatabase}

\begin{thebibliography}{10}

\bibitem{Taflove2005}
Taflove, A. and Hagness, S.~C.
\newblock {\em Computational {E}lectrodynamics: {T}he {F}inite-{D}ifference
  {T}ime-{D}omain {M}ethod.}
\newblock 3rd ed., {A}rtech {H}ouse, Norwood, {MA},  (2005).

\bibitem{Inan_Marshall2011}
Inan, U.~S. and Marshall, R.~A.
\newblock {\em Numerical electromagnetics: the {FDTD} method}.
\newblock Cambridge University Press,  (2011).

\bibitem{Cummer1997}
Cummer, S.
\newblock {\em Antennas and Propagation, IEEE Transactions on}{ \bf 45}(3),
  392--400 (1997).

\bibitem{Young2001}
Young, J. and Nelson, R.
\newblock {\em Antennas and Propagation Magazine, IEEE}{ \bf 43}(1), 61 --126
  feb.  (2001).

\bibitem{Chen1998}
Chen, Q., Katsurai, M., and Aoyagi, P.
\newblock {\em Antennas and Propagation, IEEE Transactions on}{ \bf 46}(11),
  1739 --1746 nov  (1998).

\bibitem{Georgakopoulos2002}
Georgakopoulos, S., Birtcher, C., Balanis, C., and Renaut, R.
\newblock {\em Antennas and Propagation Magazine, IEEE}{ \bf 44}(1), 134 --142
  feb  (2002).

\bibitem{Zygiridis2007}
Zygiridis, T.~T. and Tsiboukis, T.~D.
\newblock {\em Journal of Computational Physics}{ \bf 226}(2), 2372 -- 2388
  (2007).

\bibitem{Georgakopoulos2001}
Georgakopoulos, S., Renaut, R., Balanis, C., and Birtcher, C.
\newblock {\em Microwave and Wireless Components Letters, IEEE}{ \bf 11}(11),
  462--464 (2001).

\bibitem{Chen2002}
Chen, R., Wang, Z., and Chen, Y.
\newblock {\em Electronics Letters}{ \bf 38}(22), 1321--1322 (2002).

\bibitem{Hwang2006}
Hwang, K.-P. and Ihm, J.-Y.
\newblock {\em Lightwave Technology, Journal of}{ \bf 24}(2), 1048--1056
  (2006).

\bibitem{Young1996}
Young, J.
\newblock {\em Antennas and Propagation, IEEE Transactions on}{ \bf 44}(9),
  1283 --1289 sep  (1996).

\bibitem{Prokopidis2004}
Prokopidis, K.~P., Kosmidou, E.~P., and Tsiboukis, T.~D.
\newblock {\em Journal of Electromagnetic Waves and Applications}{ \bf 18}(9),
  1171--1194 (2004).

\bibitem{Li2004}
Li, J. and Chen, Y.
\newblock {\em Electronics Letters}{ \bf 40}(14), 853--855 (2004).

\bibitem{Fujii2004}
Fujii, M., Tahara, M., Sakagami, I., Freude, W., and Russer, P.
\newblock {\em Quantum Electronics, IEEE Journal of}{ \bf 40}(2), 175--182
  (2004).

\bibitem{Prokopidis2006}
Prokopidis, K.~P. and Tsiboukis, T.~D.
\newblock {\em Electromagnetic fields in mechatronics, electrical and
  electronic engineering: proceedings of ISEF '05}, chapter Higher-order
  spatial FDTD schemes for EM propagation in dispersive media,  240.
\newblock IOS Press (2006).

\bibitem{Bokil2012}
Bokil, V.~A. and Gibson, N.
\newblock {\em IMA Journal of Numerical Analysis}{ \bf 32}(3), 926--956 (2012).

\bibitem{Gedney1996}
Gedney, S.~D.
\newblock {\em Electromagnetics}{ \bf 16}(4), 399--415 (1996).

\bibitem{Fan2000}
Fan, G.-X. and Liu, Q.~H.
\newblock {\em Antennas and Propagation, IEEE Transactions on}{ \bf 48}(5), 637
  --646 may  (2000).

\bibitem{Prokopidis2008}
Prokopidis, K.~P.
\newblock {\em International Journal of Numerical Modelling: Electronic
  Networks, Devices and Fields}{ \bf 21}(6), 395--411 (2008).

\bibitem{Inan2011}
Inan, U.~S. and Go{\l}kowski, M.
\newblock {\em Principles of plasma physics for engineers and scientists}.
\newblock Cambridge University Press,  (2011).

\bibitem{Werner_Mittra_2000}
Werner, D.~H. and Mittra, R., editors.
\newblock {\em Frontiers in Electromagnetics}.
\newblock IEEE Press,  (2000).

\bibitem{Pereda2001}
Pereda, J., Vielva, L., Vegas, A., and Prieto, A.
\newblock {\em Microwave Theory and Techniques, IEEE Transactions on}{ \bf
  49}(2), 377--381 (2001).

\bibitem{Sacks1995}
Sacks, Z., Kingsland, D., Lee, R., and Lee, J.-F.
\newblock {\em Antennas and Propagation, IEEE Transactions on}{ \bf 43}(12),
  1460 --1463 dec  (1995).

\bibitem{Kuzuoglu1996}
Kuzuoglu, M. and Mittra, R.
\newblock {\em Microwave and Guided Wave Letters, IEEE}{ \bf 6}(12), 447 --449
  dec  (1996).

\bibitem{Yefet2001}
Yefet, A. and Petropoulos, P.~G.
\newblock {\em Journal of Computational Physics}{ \bf 168}(2), 286--315 (2001).

\bibitem{Balanis1997}
Balanis, C.~A.
\newblock {\em Antenna Theory: Analysis and Design}.
\newblock Wiley, 2nd edition,  (1997).

\end{thebibliography}

\end{document}